\documentclass[prb,twocolumn,showpacs,superscriptaddress,floatfix]{revtex4}  

\usepackage{graphicx}
\usepackage[dvips]{epsfig}
\usepackage{dcolumn}
\usepackage{bm}
\usepackage[mathlines]{lineno}
\usepackage{mathtools}

\begin{document}

\preprint{APS/123-QED}

\title{Spin-Wave Propagation in the Presence of Inhomogeneous Dzyaloshinskii-Moriya Interaction}

\author{Seung-Jae Lee}
\affiliation{KU-KIST Graduate School of Converging Science and Technology, Korea University, Seoul 02841, Korea}

\author{Jung-Hwan Moon}
\affiliation{Department of Materials Science and Engineering, Korea University, Seoul 02841, Korea}

\author{Hyun-Woo Lee}
\affiliation{PCTP and Department of Physics, Pohang University of Science and Technology, Pohang 37673, Korea}

\author{Kyung-Jin Lee}
\email{kj_lee@korea.ac.kr}
\affiliation{KU-KIST Graduate School of Converging Science and Technology, Korea University, Seoul 02841, Korea}
\affiliation{Department of Materials Science and Engineering, Korea University, Seoul 02841, Korea}

\date{\today}

\begin{abstract}
We theoretically investigate spin-wave propagation through a magnetic metamaterial with spatially modulated Dzyaloshinskii-Moriya interaction.
We establish an effective Sch{\"o}dinger equation for spin-waves and derive boundary conditions for spin-waves passing through the boundary between two regions having different Dzyaloshinskii-Moriya interactions. Based on these boundary conditions, we find that the spin-wave can be amplified at the boundary and the spin-wave bandgap is tunable either by an external magnetic field
or the strength of Dzyaloshinskii-Moriya interaction, which offers a spin-wave analogue of the field-effect transistor in traditional electronics. 

\end{abstract}


\maketitle



\section{\label{sec:level1}Introduction}
The Dzyaloshinskii-Moriya interaction (DMI) is the antisymmetric component of quantum mechanical exchange interaction in magnetic systems~\cite{Dzyaloshinskii,Moriya}. Three prerequisites of the DMI are the exchange interaction, the spin-orbit interaction, and the inversion symmetry breaking. All these three prerequisites are simultaneously satisfied in material systems such as B20 structures~\cite{Rossler2006,Uchida,Huang} and ferromagnet/heavy metal bilayer structures~\cite{Fert1980,Bode2007,Heide2008,Udvardi2009,Zakeri2010,Costa,Kim}. The DMI affects the equilibrium spin texture and consequently magnetization dynamics by stabilizing chiral domain walls~\cite{Thiaville,Chen2013,Beach,Parkin} or magnetic skyrmions~\cite{Muhlbauer2009,Yu2010Nat,Jiang2015,Woo,Moreau,Boulle,Yang2017}. The DMI also causes the nonreciprocal spin-wave propagation~\cite{Zakeri,Landeros,Moon}, which is widely used to estimate the strength of DMI~\cite{Di,Nembach,Cho,Lee2016,Seki}. Recent works found that the DMI effect on the spin wave propagation can also result in unidirectional caustic beams~\cite{JV}, spin wave diodes~\cite{Lan}, and spin wave fibers~\cite{Yu2016}, opening rich spin wave physics and wide applications in functional devices based on spin waves.

The magnonic crystal is a magnetic metamaterial with alternating magnetic properties that serve as periodic potential for spin waves passing through it~\cite{Nikitov,Serga,Lee2009,Lenk,Ding,Barsukov,Tacchi2011,Karenowska,Duerr,Chumak2012,Tacchi2012,Krawczyk,Chumak2015,Mruczkiewicz}. As spin waves, the collective precessional motion of localized electron spins, do not involve the motion of electrons, magnonic devices avoid Joule heating and thus allow low-power computing~\cite{Serga,Lenk,Krawczyk,Chumak2015}. Moreover, their wave properties provide distinct functionalities~\cite{Vogt,Vogel,Adeyeye,Kwon,Sekiguchi} such as multi-input/output (non-linear) operations~\cite{Khitun2012,Khitun2013}. Despite their attractive features, however, magnonic devices suffer from a small on/off ratio of spin-wave signal. We note that in traditional electronic logic devices based on field-effect transistors, the source-drain current substantially varies by a gate voltage. For practical use of magnonic devices, therefore, it is of critical importance to largely modulate spin-wave signals for a given spin-wave frequency. For this purpose, a possible way is to modulate the spin-wave bandgap by an external means; for a given spin-wave frequency, opening/closing the bandgap at the frequency offers a large change in the signal of propagating spin-waves through a magnonic crystal.
In this work, we theoretically demonstrate that magnonic crystals with alternating DMI show an efficient modulation of spin-wave signal. We consider the interfacial
DMI present in ferromagnet/heavy metal bilayers but the working principle is also applicable to arbitrary DMI symmetries, including the bulk DMI in B20 structures, by rotating the DM vector.

We first focus on a magnetic thin film with a vanishing demagnetization effect, which is experimentally achievable by tuning the thickness of a thin film having the surface perpendicular anisotropy. In such thin film structures, a spatial modulation of interfacial DMI can be realized by a local gating~\cite{Nawaoka}, a local modulation of the interface between ferromagnetic layer and heavy metal layer~\cite{Schmid,Torrejon,Manchon2016}, or a local variation of the heavy metal thickness~\cite{Tacchi2017}. For a magnonic crystal with spatially modulated DMI, we establish an effective Sch\"{o}dinger equation for spin-waves and derive spin-wave boundary conditions at the boundary between two regions having different DMI values. With these boundary conditions, we construct a spin-wave version of the Kronig-Penny model~\cite{KP}, which is a simplified model for an electron in a one-dimensional periodic potential. At the end of this paper, we show that our finding, an efficient tunablility of spin-wave bandgap in DMI-modulated magnonic crystals also holds for magnetic thin films with a finite demagnetization effect.

\section{\label{sec:level2}ANALITYCAL AND NUMERICAL RESULTS}
\subsection{\label{sec:level3}BOUNDARY CONDITION}
Let us consider a one-dimensional magnetic thin film where the magnetization ${\bf m}$ is allowed to vary in $x$ direction. In this system, the magnetic energy density $W$ reads
\begin{eqnarray}\label{W}
W = A \left(\partial_x {\bf m}\right)^2-D(x) {\bf m} \cdot \left(\hat{\bf y} \times \partial_x {\bf m}\right) - M_S {\bf m}\cdot {\bf H},
\end{eqnarray}
where $\partial_x \equiv \partial/\partial x$, $A$ is the symmetric exchange constant, $D(x)$ is the DMI constant that is spatially inhomogeneous along the $x$-direction, $\hat{\bf y}$ is the unit vector along the perpendicular to both spin-wave propagation direction (i.e., $\hat{\bf x}$) and the thickness direction $\hat{\bf z}$, $M_{\rm S}$ is the saturation magnetization, and ${\bf H}$ is the external magnetic field applied in the film plane. Here we assume that the only DMI is inhomogeneous by interface engineering. The corresponding equation of motion is the Landau-Lifshitz-Gilbert (LLG) equation, given as
\begin{equation}\label{LLG}
\frac{\partial {\bf m}}{\partial t}=\frac{\gamma}{M_{\rm S}} {\bf m} \times \frac{\delta W}{\delta {\bf m}}+\alpha {\bf m} \times \frac{\partial {\bf m}}{\partial t},
\end{equation}
where $\alpha$ is the damping, $\bf {m}$ = $\bf {m}_{\rm 0}$ + $\delta\bf {m}$, $\bf {m}_{\rm 0}$ is the position-independent magnetization, $\delta\bf {m}$ = $(0,s_{\rm \theta},s_{\rm \phi})$ is the spin-wave contribution in the spherical coordinate $(s_\theta^2 \ll 1, s_\phi^2 \ll 1)$, and $\theta$ and $\phi$ are the polar and azimuthal angles, respectively.

Neglection the damping, we obtain the effective one-dimensional Schr{\"o}dinger equation for spin-wave wave-function $\psi(=s_\theta + is_\phi)$ from the LLG equation as
\begin{eqnarray}\label{Sch}
i\hbar \frac{\partial \psi}{\partial t} = \hat {\cal H} \psi &=& \left[\frac{\hat {p}_x^2}{2 m^*} -\frac{(\bf {m}_{\rm 0} \cdot \hat{\bf y})}\hbar  \alpha_{\rm D}(x) \hat {p}_x \right. \\ \nonumber
&&\left.-\frac{(\bf {m}_{\rm 0} \cdot \hat{\bf y})}{2i}\frac{\partial \alpha_{\rm D}(x)}{\partial x}+\gamma \hbar {\mu_0}{H}\right]\psi ,
\end{eqnarray}
where $\hat {p}_x (\equiv -i \hbar \partial/\partial x)$ is the momentum operator, $m^* (\equiv \hbar M_{\rm S}/4 \gamma A)$ is the effective spin wave mass, ${\alpha_{\rm D}(x)}/{\hbar} (\equiv 2\gamma D(x)/M_{\rm S})$ is the DM velocity for spin-waves, and $\gamma$ is the gyromagnetic ratio. We note that the third term in the bracket of Eq.~(\ref{Sch}) is essential for $\hat {\cal H}$ to be a Hermitian operator. For a system with homogeneous DMI (i.e., $\partial \alpha_{\rm D}(x)/\partial x=0$) and in the absence of the external field, Eq.~(\ref{Sch}) reproduces our previous result~\cite{Manchon}.

From the continuity of wave function and integration Eq.~(\ref{Sch}) for $x$, we obtain the boundary conditions for spin-wave wavefunction $\psi$ at the boundary between two regions (i.e., $D = D_1$ for $x < 0$ and $D = D_2$ for $x \ge 0$) as 

\begin{eqnarray}
\psi_1(x=0) &=& \psi_2(x=0) \label{BC1} \\
\left.\frac{d\psi_1}{dx}\right|_{x=0}-\left.\frac{d\psi_2}{dx}\right|_{x=0} &=& i \frac{\Delta D}{2A} ({\bf {m}_{\rm 0}} \cdot \hat{\bf y}) \psi(x=0) \label{BC2},
\end{eqnarray}
where $\Delta D = D_2-D_1$. We note that the second boundary condition (Eq.~(\ref{BC2})) describes the effect of a DMI step on spin waves, originating from the gradient of DMI in ${\cal H}$.

In order to verify the boundary conditions to the spin-wave propagation through a DMI step, we consider a plane spin-wave and ${\bf m_{\rm 0}}=\hat{\bf y}$, the incident ($\psi_I$) and reflected ($\psi_R$) waves in the region 1 where $D=D_1$, and the transmitted ($\psi_T$) wave in the region 2 where $D=D_2$ are given as
\begin{equation}\label{amp1}
\psi_I = I {\rm e}^{ik_1x}, \quad \psi_R = A_0 {\rm e}^{-ik_2x}, \quad \psi_T = B_0 {\rm e}^{ik_3x}
\end{equation}
where $I$, $A_0$, and $B_0$ ($k_1$, $k_2$, and $k_3$) are the spin wave amplitudes (wavenumbers) of incident, reflected, and transmitted waves, respectively. From the spin wave dispersion in each region,
\begin{equation}\label{dis}
\omega=\gamma \mu_0 \left(Jk^2-D_i^*k+H\right)
\end{equation}
where $J=2A/\mu_0 M_{\rm S}$ and $D_i^* = 2D_i/\mu_0 M_{\rm S}$ of the $i$th region ($i=1,2$), combined with the boundary conditions, we obtain $A_0/I$ and $B_0/I$, given as

\begin{figure}[t]
  \centering
  \includegraphics[scale=1.]{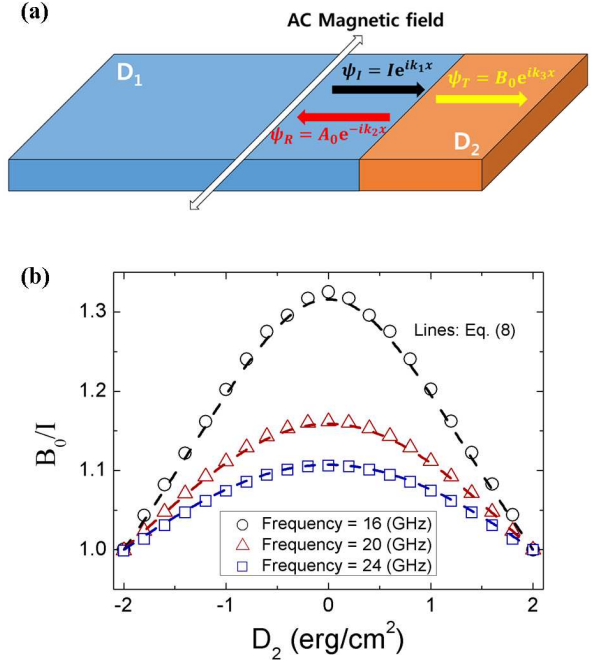}
  \caption{(color online) (a) The coordinate system and schematic illustration of spin-wave transmission and reflection at a DMI step. (b) The ratio of transmitted spin-wave ($B_0$) to incident spin-wave ($I$) as a function of $D_2$. Parameters: The saturation magnetization $M_s$ = 800 kA/m, the exchange stiffness $A = 1.3\times10^{-11}$ J/m, the DMI of the region 1 $D_1$ = 2 mJ/m$^2$, the external field $H$ applied along the $y$-axis = 0.5 T, the Gilbert damping $\alpha$ = 0.01, the unit cell size along the spin-wave propagation direction = 2 nm. \\}%
  \label{fig:FIG1}
\end{figure}

\begin{eqnarray}
\frac{A_0}{I} &=& \frac{\sqrt{(D_1^*)^2+4JH^*}-\sqrt{(D_2^*)^2+4JH^*}}{\sqrt{(D_1^*)^2+4JH^*}+\sqrt{(D_2^*)^2+4JH^*}}, \label{AI}\\
\frac{B_0}{I} &=& \frac{2\sqrt{(D_1^*)^2+4JH^*}}{\sqrt{(D_1^*)^2+4JH^*}+\sqrt{(D_2^*)^2+4JH^*}}, \label{BI}
\end{eqnarray}
where $H^*=\omega/\gamma\mu_0-H$. One finds from Eqs.~(\ref{AI}) and (\ref{BI}) that the current conservation holds, i.e., $R+T\equiv 1$ where $R=|A_0 \sqrt{|v_R/v_I|}/I|^2$ and $T=|B_0 \sqrt{|v_T/v_I|}/I|^2$. Here, $v_I$, $v_R$, and $v_T$ are the group velocities of incident, reflected, and transmitted waves at $x \rightarrow 0$, respectively. The current conservation justifies the boundary conditions. Moreover, Eq.~(\ref{BI}) shows $B_0/I>1$ when $|D_1| > |D_2|$, i.e., the spin-wave amplification at a DMI step. Numerical simulations based on the LLG equation quantitatively reproduces Eq.~(\ref{BI}) [Fig. \ref{fig:FIG1}], also justifying the validity of the boundary conditions.
\subsection{\label{sec:level5}KRONIG-PENNY MODEL}
We next establish a spin-wave version of the Kronig-Penny model. We consider a one-dimensional magnonic crystal with a periodic DMI modulation;

\begin{eqnarray}\label{MC}
{\rm Region \quad 1}: D(x) &=& D_1, \quad na < x \le \left(n+1/2\right)a \\ \nonumber
{\rm Region \quad 2}: D(x) &=& D_2, \quad \left(n+1/2\right)a < x \le (n+1)a,
\end{eqnarray}
where $n=0,1,2,...,$ and $a/2$ is the width of a homogeneous DMI region. Based on the boundary conditions [Eqs.~(\ref{BC1}-\ref{BC2})] and the Bloch's theorem, we obtain an equation for a spin-wave version of the Kronig-Penny model as

\begin{eqnarray}\label{KP1}
\cos \left(ka+\frac{D_1^*+D_2^*}{4J}a\sin\theta \right) &&= \cos \frac{a\mu}{2} \cos \frac{a\nu}{2} \\ \nonumber
&&-\frac{\mu^2+\nu^2}{2\mu\nu}\sin \frac{a\mu}{2} \sin \frac{a\nu}{2},
\end{eqnarray}

where
\begin{eqnarray}\label{munu}
\mu = \sqrt{(D_1^*\sin\theta/2J)^2+H^*/J},\\ \nonumber \quad \nu = \sqrt{(D_2^*\sin\theta/2J)^2+H^*/J}.
\end{eqnarray}
Here $\theta$ is the angle between the magnetization $\bf{m}_0$ and the spin-wave propagation direction $\hat{\bf x}$. Equation~(\ref{KP1}) allow us to identify the necessary conditions for finite spin-wave bandgaps. When $\theta=0$ (i.e., $\bf{m}_0$ is aligned in the $x$-axis), $\mu$ and $\nu$ are identical so that the right-hand-side of Eq.~(\ref{KP1}) becomes $\cos\left(\mu\nu\right)$; i.e., no spin-wave bandgap is expected except for the first forbidden spin-wave band that originates from the external field $H$ and ranges from zero to a finite spin-wave frequency. When $\theta\ne0$ and $|D_1|\ne|D_2|$, in addition to the first forbidden band, there are always multiple frequency-ranges in which the absolute value of the right-hand-side of Eq.~(\ref{KP1}) is greater than the unity. These frequency-ranges correspond to the additional spin-wave bandgaps.

\subsection{\label{sec:level6}TUNABILITY OF SPIN-WAVE BANDGAPS}
\begin{figure}[t]
  \centering
  \includegraphics[scale=1.4]{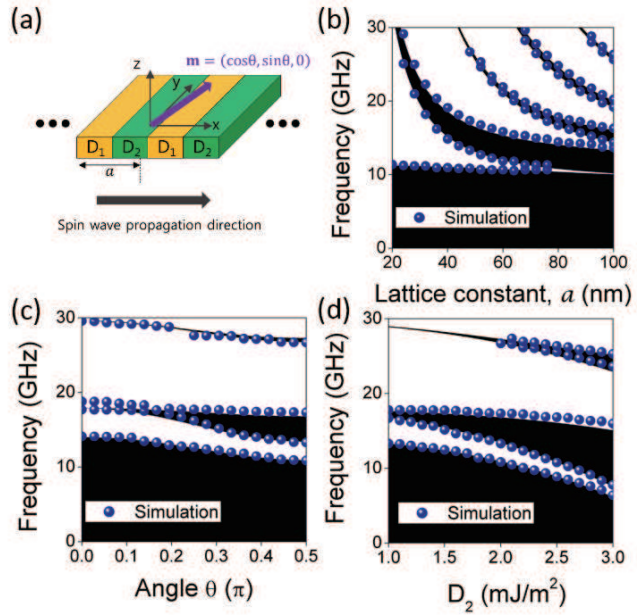}
  \caption{(color online) (a) The coordinate system and schematic illustration of a magnonic crystal with alternating DMI. (b) The spin-wave frequency versus the lattice constant a of an alternating DMI region ($\theta = \pi / 2$, $D_1$ = 0 mJ/m$^2$, and $D_2$ = 2 mJ/m$^2$). (c) The spin-wave frequency versus the angle $\theta$ ($a$ = 48 nm, $D_1$ = 0 mJ/m$^2$, and $D_2$ = 2 mJ/m$^2$). (d) The spin-wave frequency versus $D_2$ ($a$ = 48 nm, $\theta=\pi/2$, and $D_1$ = 0 mJ/m$^2$). In (b)-(d), black (white) regions are spin-wave forbidden (allowed) bands calculated from the spin-wave version of the Kronig-Penny model. Circular symbols in (b)-(d) correspond to numerical simulation results. Unless specified, the parameters used for the calculations are the same as in Fig.~\ref{fig:FIG1}. \\}%
  \label{fig:FIG2}
\end{figure}

In Fig.~\ref{fig:FIG2}, we summarize the spin-wave forbidden and allowed bands for various parameters. Key features of the spin-wave bands are as follows. The allowed bands become narrower with increasing the lattice constant $a$ [Fig.~\ref{fig:FIG2}(b)], like results for magnonic crystals constructed with patterned defect structure~\cite{Barsukov}, because the width of the DMI-induced potential barrier for spin-waves increases. A similar narrowing of the allowed bands occurs as the angle $\theta$ increases from 0 to $\pi/2$ [Fig.~\ref{fig:FIG2}(c)], because the DMI contribution to the spin-wave energy is proportional to $\sin\theta$. For a fixed $D_1$, a larger $|\Delta D|$ also results in narrower allowed bands [Fig.~\ref{fig:FIG2}(d)], because the height of potential barrier for spin-waves increases. The angle variation shown in Fig.~\ref{fig:FIG2}(c) can be realized by rotating the equilibrium magnetization $\bf{m}_0$ by means of an external magnetic field. The injection of an in-plane current should also work for this purpose as it generates spin-orbit torque in ferromagnet/heavy metal bilayers~\cite{Miron2011,Liu}. The DMI variation shown in Fig.~\ref{fig:FIG2}(d) can be also realized by a local gating~\cite{Nawaoka} that modifies the DMI locally. We note that in all cases, the change in the spin-wave band structure can be very large; e.g., the width of the second forbidden band in Fig.~\ref{fig:FIG2}(c) varies from $\approx$ 0 GHz at $\theta = 0$ to $≈$ 5 GHz at $\theta = \pi/2$. We note that such a large change in the bandgap is obtained not only for an abrupt variation of DMI (Fig.~\ref{fig:FIG2}), but also for a much smoother sinusoidal variation of DMI (not shown). Therefore, one enhances the on/off ratio of spin-wave signals substantially by locating the spin-wave frequency in the frequency ranges in which the bandgap varies with $\theta$ or $|\Delta D|$. This efficient tunability of the spin-wave bandgap in DMI-modulated magnonic crystals is able to mimic the field-effect transistors in traditional electronics.

In Fig.~\ref{fig:FIG2}, we also compare the boundaries between the forbidden and allowed bands, obtained from the spin-wave version of the Kronig-Penny model (Eq.(~\ref{KP1})), with those obtained by numerically solving the LLG equation (circular symbols). They are in reasonable agreement except for deviations in spin-wave bands as the angle $\theta$ [Fig.~\ref{fig:FIG2}(c)]. The reason of these deviations is as follows. In the presence of DMI step at $x=i_0$($D=D_1$ for $x < i_0$ and $D=D_2$ for $x \geq i_0$), the DM energy ${\bf{E}}_{{\rm {DM}}, i_0}$ is given by
\begin{equation}\label{EDM_re}
\frac{{\bf{E}}_{{\rm {DM}}, i_0}}{{2\delta x}}=D_1\hat{\bf y}\cdot \left({\bf {m}}_{i_0-1}\times {\bf {m}}_{i_0} \right)+D_2\hat{\bf y}\cdot \left({\bf {m}}_{i_0}\times {\bf {m}}_{i_0+1} \right)
\end{equation} 
here we substitute $D_1$ and $D_2$ for $\bar{D}$ and $\Delta D$ ($D_1=\bar{D}-\Delta D/2$ and $D_2=\bar{D}+\Delta D/2$), then we obtain follow general equations for the DM energy $E_{\rm DM}$ and the corresponding DM field $\bf{H}_{\rm DM}$ at the step. 
\begin{equation}\label{EDM}
{E}_{\rm DM} = -\bar{D} {\bf m} \cdot \left(\hat{\bf y}\times\frac{\partial \bf m}{\partial x}\right)-\frac{\Delta{D}}{2\delta x} {\bf m}\cdot \left(\hat{\bf y} \times \bar{\bf m}\right),
\end{equation}
\begin{equation}\label{HDM}
{\bf{H}}_{\rm DM} = \frac{2}{\mu_0 M_{\rm S}} \left({\bar{D}}\hat{\bf y}\times\frac{\partial \bf m}{\partial x}+\frac{\Delta D}{2\delta x} \hat{\bf y}\times {\bar{\bf m}}\right),
\end{equation}
where $\bar{D}$ $(\bar{\bf m})$ is the average DMI constant (magnetization) of the nearest neighbor sites acrossing the DM step, and $\delta x$ is the lattice constant for $\bf {m}$. The second term on the right-hand-side of Eq.~(\ref{HDM}) is an addtional effective field originating from the DMI step and acts like a magnetic field applied along the $z$-axis at $\theta = 0$ because $\bar{\bf m}\approx \bf m_{\rm 0} =\hat{\bf x}$. When $\bf m_{\rm 0}$ deviates from the $y$-axis, this additional effective field tilts the magnetization at the DMI step, causing a deviation of the magnetic state from the uniform state. Therefore the micromagnetic simulation results has a deviation near $\theta=0$, where the forbidden band is not expected in Eq.~(\ref{KP1}).  As shown in Fig.~\ref{fig:FIG2}, however, the effect of this additional tilting on the spin-wave band structure is rather weak and does not alter our main conclusion for tunability of the spin-wave bandgap.

\begin{figure}[t]
  \centering
  \includegraphics[scale=0.28]{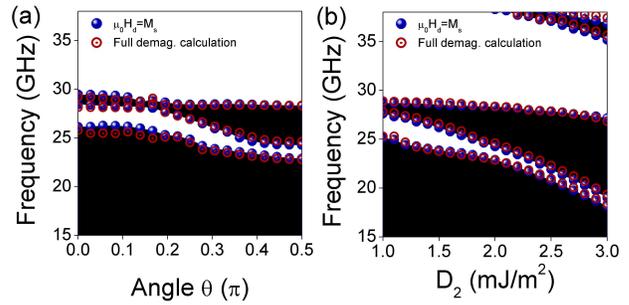}
  \caption{(color online) The spin-wave forbidden and allowed bands in the presence of the demagnetization field $\mu_0 H_d$. (a) The spin-wave frequency versus the angle $\theta$ ($a$ = 48 nm, $D_1$ = 0 mJ/m$^2$, and $D_2$ = 2 mJ/m$^2$). (b) The spin-wave frequency versus $D_2$ ($a$ = 48 nm, $\theta=\pi/2$, and $D_1$ = 0 mJ/m$^2$). In (a) and (b), blue symbols and black (white) regions are spin-wave forbidden (allowed) bands calculated by micromagnetic simulations with $\mu_0 H_d (= M_{\rm S})$. Red symbols in (a) and (b) correspond to simulation results for including full magnetostatic interaction. Unless specified, the parameters used for the calculations are the same as in Fig.~\ref{fig:FIG1}. \\}%
  \label{fig:FIG3}
\end{figure}

Finally we show that the efficient tunability of the spin-wave bandgap is realized even with a finite demagnetization effect. In the presence of the demagnetization effect, the magnetization undergoes an elliptical precession so that one cannot convert the LLG equation to a Schr{\"o}dinger-like equation. For this case, therefore, we obtain variations of the spin-wave band structure by numerically solving the LLG equation [Fig.~\ref{fig:FIG3}]. We obtain a qualitatively similar trend to the case with no demagnetization effect [Fig.~\ref{fig:FIG2}]; the allowed band width decreases as either $\theta$ or $|\Delta D|$ (or $D_2$ for a fixed $D_1$) increases. This result confirms that the spin-wave bandgap in DMI-modulated magnonic crystals is efficiently tunable regardless of the demagnetization effect. For a comparison, we also plot results for including full magnetostatic interaction (for that purpose, we use the thin film in Fig.~\ref{fig:FIG1}, discretized along the spin-wave propagation direction. Here the length of the thin film is 4 $\mu$m , the width is 400 nm, and the thickness is 1.5 nm). The small deviation of spin-wave bandgap in both results is observed but it could be disregarded as anticipated because the film is sufficiently thin.

\section{SUMMARY}
In conclusion, we propose that magnonic crystals with spatially modulated DMI are highly efficient to change the spin-wave bandgap by an external means such as a magnetic field, an in-plane current, and a perpendicular voltage gating. This high efficiency is caused by the fact that the DMI contribution to the spin-wave energy is sizable and highly anisotropic depending on the relative orientation of the equilibrium magnetization with respect to the spin-wave propagation direction. We note that the nonlocal magnetostatic interaction, which we ignore in this work, also contributes to the anisotropic dispersion as the spin-wave dispersion of the backward volume mode (i.e., $\bf m_{\rm 0}\parallel \bf k$) is different from that of the surface mode (i.e., $\bf m_{\rm 0}\perp \bf k$). For an experimentally accessible $D$, however, the DMI contribution is much stronger than the contribution from the magnetostatic interaction~\cite{Moon}. Moreover, the magnetostatic contribution is effective in the small $k$-limit, which makes the scaling of magnonic devices difficult. In contrast, the DMI works for an intermediate to a large $k$ so that magnonic crystals with DMI modulation are expected to be more suitable for higher density devices. The large tunability of spin-wave bandgap can enhance the on/off ratio of spin-wave signals as spin-waves are unable to propagate when the bandgap is large. The proposed magnonic crystals may be useful to reproduce various functionalities of field-effect transistors in traditional electronics at low-power consumption. We end this paper by noting that even though we focus on spin-wave dynamics, our finding will be useful to understand domain wall or skyrmion dynamics in the presence of inhomogeneous DMI as magnetic solitons can be described by spin-wave packets.\\
\begin{acknowledgments}
This work was supported by the National Research Foundation of Korea (NRF-2015M3D1A1070465, NRF-2017R1A2B2006119, NRF-2011-0030046) and KU-KIST Graduate School of Converging Science and Technology Program.
\end{acknowledgments}

\end{document}